\documentclass{emulateapj}
\shorttitle{Dust in SN~2010jl}
\shortauthors{Smith et al.}
\begin{document}

\title{Systematic Blueshift of Line Profiles in the
  Type~II\lowercase{n} Supernova 2010\lowercase{jl}: Evidence for
  Post-Shock Dust Formation?}

\author{Nathan Smith\altaffilmark{1}, Jeffrey M.\
  Silverman\altaffilmark{2}, Alexei V.\ Filippenko\altaffilmark{2}
  Michael C.\ Cooper\altaffilmark{3}, Thomas Matheson\altaffilmark{4},
  Fuyan Bian\altaffilmark{1}, Benjamin J.\ Weiner\altaffilmark{1}, and
  Julia M. Comerford\altaffilmark{5}}
\altaffiltext{1}{Steward Observatory, University of Arizona, 933 North
  Cherry Avenue, Tucson, AZ 85721; Email:
  nathans@as.arizona.edu.} 
\altaffiltext{2}{Department of Astronomy, University of California,
  Berkeley, CA 94720-3411.}
\altaffiltext{3}{Department of Physics and Astronomy, University of
  California, 4129 Frederick Reines Hall, Irvine, CA 92697-4575.}
\altaffiltext{4}{National Optical Astronomy Observatory, 950 North
  Cherry Avenue, Tucson, AZ 85719-4933.} 
\altaffiltext{5}{Astronomy Department, University of Texas at Austin,
  Austin, TX 78712}

\begin{abstract}

  Type~IIn supernovae (SNe) show spectral evidence for strong
  interaction between their blast wave and dense circumstellar
  material (CSM) around the progenitor star.  SN~2010jl was the
  brightest core-collapse supernova in 2010, and it was a Type~IIn explosion
  with strong CSM interaction.  Andrews et al.\ recently reported
  evidence for an infrared (IR) excess in SN~2010jl, indicating either
  new dust formation or the heating of CSM dust in an IR echo.  Here
  we report multi-epoch spectra of SN~2010jl that reveal the tell-tale
  signature of new dust formation: emission-line profiles becoming
  systematically more blueshifted as the red side of the line is
  blocked by increasing extinction.  The effect is seen clearly in the
  intermediate-width (400--4000 km s$^{-1}$) component of H$\alpha$
  beginning roughly 30 days after explosion.  Moreover, we present
  near-IR spectra demonstrating that the asymmetry in the hydrogen-line 
  profiles is wavelength dependent, appearing more pronounced at
  shorter wavelengths.  This evidence suggests that new dust grains
  had formed quickly in the post-shock shell of SN~2010jl arising from
  CSM interaction.  Since the observed dust temperature has been
  attributed to an IR echo and not to new dust, either (1) IR excess
  emission at $\lambda \, < \, 5 \, \mu$m is not a particularly
  sensitive tracer of new dust formation in SNe, or (2) some
  assumptions about expected dust temperatures might require further
  study.  Lastly, we discuss one possible mechanism other than dust
  that might lead to increasingly blueshifted line profiles in
  SNe~IIn, although the wavelength dependence of the asymmetry argues
  against this hypothesis in the case of SN~2010jl.

\end{abstract}


\section{INTRODUCTION}

The formation of dust grains in core-collapse supernova (SN)
explosions is of interest in part because supernovae (SNe)
contributed to the dust budget of the early interstellar medium seen
in high-redshift galaxies.  SNe that may arise from very massive stars
are of particular importance in this regard (e.g., Cherchneff \& Dwek
2010).

New dust formation in a SN may give rise to three observable
signatures seen concurrently: (1) infrared (IR) excess due to thermal
emission from hot or warm dust, (2) increased rate of fading of the
optical flux (typically on the radioactive decay tail), and (3)
progressive and systematic blueshift of emission-line profiles when
the receding parts of the ejecta are increasingly blocked by new dust.
The first two might also arise from an IR echo due to pre-existing
dust in the circumstellar material (CSM) or from the escape of
radioactive luminosity due to decreasing optical depth in the SN
ejecta, respectively, but the presence of all three concurrently is
generally taken to imply that these effects are caused by the 
formation of new dust grains.  Clear observational evidence of all
three effects is rare, however, limited to cases of very nearby SNe.
In addition to the first well-established case of SN~1987A (Danziger
et al.\ 1989; Lucy et al.\ 1989; Gehrz \& Ney 1989; Wooden et al.\
1993; Colgan et al.\ 1994), only a few other compelling examples of
dust formation in normal SN ejecta have been found.  Thus far, though, 
dust masses estimated for normal core-collapse SNe ($M_{\rm ZAMS} < 
20~ {\rm M}_{\odot}$ progenitor stars) are very low, being
1--3 orders of magnitude too small to account for the dust budget
inferred in high-redshift galaxies (see, e.g., Elmhamdi et al.\ 2003;
Sugerman et al.\ 2006; Meikle et al.\ 2007; Kotak et al.\ 2009;
Andrews et al.\ 2010; Szalai et al.\ 2011; Meikle et al.\ 2011).  In
these SNe, the dust formation occurs in the rapidly expanding SN
ejecta a few hundred days after explosion.  


\begin{figure*}
\epsscale{0.8}
\plotone{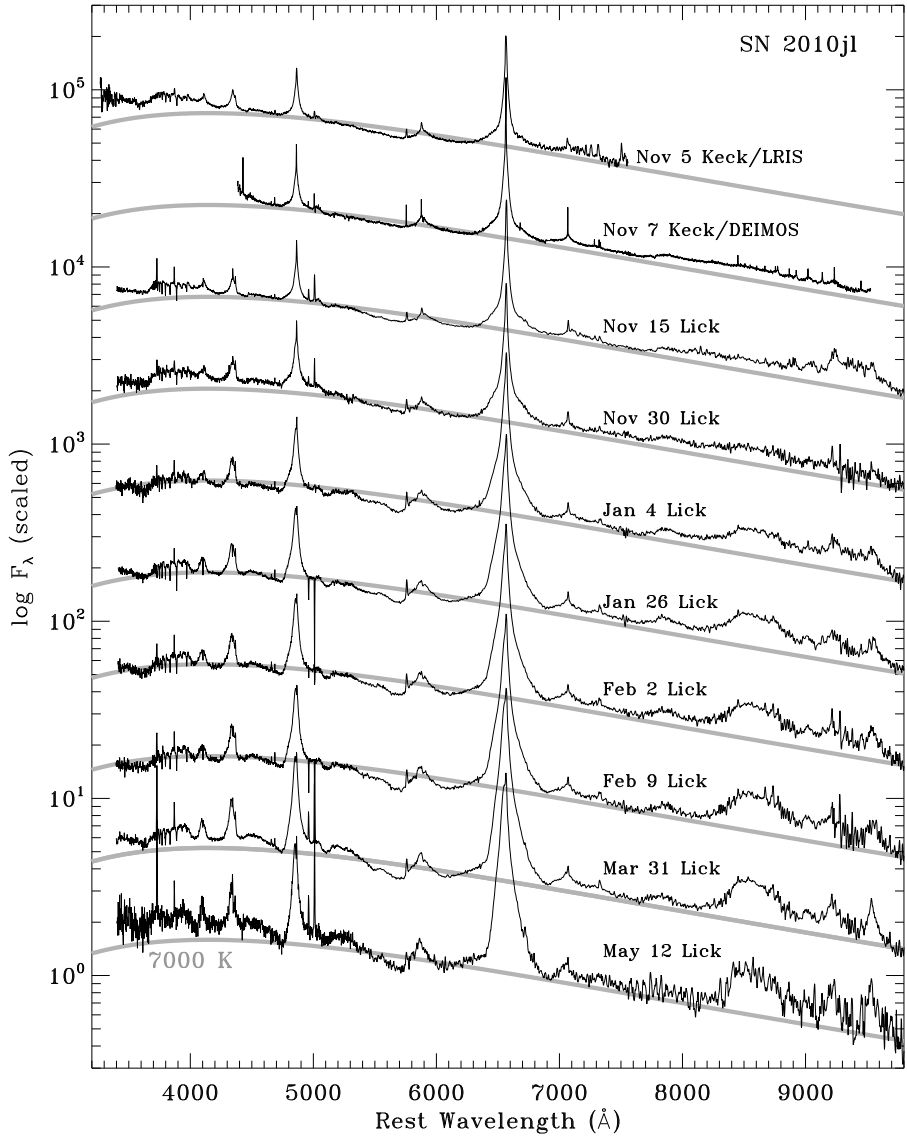}
\caption{Optical spectra of SN~2010jl showing the spectral evolution
  from early times a few days post-discovery until 6 months later (see
  Table 1).  Overall, the spectrum appears constant during this time
  period, with roughly the same $\sim$7,000 K characteristic continuum
  temperature at these wavelengths.  An important devlopment, however,
  is the appearance of the Ca~{\sc ii} IR triplet.}
\label{fig:spec}
\end{figure*}

Recent far-IR observations of
SN~1987A have revealed an unexpectedly large amount of very cold dust
in the SN ejecta (Matsuura et al.\ 2011).  The $\sim$20 K dust is much
cooler than expected, and the huge mass of 0.4--0.7 M$_{\odot}$ is
much more than the 10$^{-4}$ to 10$^{-3}$ M$_{\odot}$ that had been
inferred based on IR emission observed a few hundred days after the
explosion (Wooden et al.\ 1993).  It should be noted that the early
mass estimates around 10$^{-4}$ M$_{\odot}$ were lower limits due to
the fact that the dust may have resided in optically thick clumps, but
the large mass detected recently in the far-IR nevertheless presents a
significant mystery, and implies that vast amounts of dust may somehow
escape detection by IR emission.


Recent evidence has been accumulating that a different mode of dust
formation is also present in some SNe, where dust grains can form
efficiently and very {\it early} in the cold dense shell arising from
strong CSM interaction.  In this scenario, the high density in the
radiative post-shock layers allows the post-shock gas to cool rapidly,
thereby permitting dust formation.  The first clear evidence for this
mode of dust formation was seen in SN~2006jc, which showed all three
effects described above.  In addition to a near-IR excess from very hot
dust and an increased fading rate, it showed a systematic blueshift in
He~{\sc i} emission lines formed in the CSM interaction region (Smith
et al.\ 2008a).  In particular, Smith et al.\ (2008a) demonstrated
that the relative strength of the systematic blueshift seen in He~{\sc
  i} lines at different wavelengths was consistent with the
wavelength-dependent extinction one expects from small dust grains.
This provided the first solid example of dust formed in the dense
post-shock cooling zone of a SN dominated by CSM interaction, although
it had been noted earlier as a possibility in the case of the SN~IIn
1998S (Pozzo et al.\ 2004).  Subsequently, Matilla et al. (2008)
presented mid-IR data for SN~2006jc and additional spectra, yielding
similar results.

SN~2006jc was a Type~Ibn SN, and it is possible that C-rich SN ejecta
crossing the reverse shock may have aided the efficiency of dust
formation (see Smith et al.\ 2008a; Matilla et al.\ 2008).  However,
evidence for post-shock dust formation has also been reported in a few
examples of Type~IIn SNe that are not expected to have C-rich ejecta.
These include SN~1998S (Pozzo et al.\ 2004), SN~2005ip (Smith et al.\
2009; Fox et al.\ 2009), and SN~2006tf (Smith et al.\ 2008b).  More
recently, some of the SNe~IIn studied by Fox et al.\ (2011) also show
possible evidence for dust formation, although these authors conclude
based on the observed dust temperature that the bulk of the IR excess was
probably due to an IR echo in most cases.  Fox et al.\ (2011) provide
a more detailed discussion of the issue of dust formation and IR
echoes in SNe~IIn.  SNe~IIn constitute only about 6--9\% of
core-collapse SNe (Smith et al.\ 2011b; Li et al.\ 2011), so even if
they are very efficient dust producers, it is not clear that they can
account for the dust present in high-redshift galaxies unless they are
more common at high redshift.


Andrews et al.\ (2011) recently reported an IR excess in the
very bright Type IIn SN~2010jl.  They note that the IR
excess may arise from either newly formed dust or an IR echo from
pre-existing dust, although they favor the hypothesis of an IR echo
because of the observed dust temperature.  In this paper we consider
the evolution of optical emission-line profiles in SN~2010jl, which
suggest the presence of newly formed dust as well.

Supernova (SN)~2010jl was discovered on 2010 Nov.\ 3.52 (UT dates are
used throughout this paper) by Newton \& Puckett (2010).  It
was intrinsically luminous, with a peak absolute magnitude of about
$-$20 (Smith et al.\ 2011a).  Early-time spectra showed that it is a
Type~IIn SN (Benetti et al.\ 2010; see also Filippenko 1997).  In a
previous paper (Smith et al.\ 2011a), we analyzed pre-explosion
archival images of the field of SN~2010jl obtained with the {\it
  Hubble Space Telescope (HST)}, showing that the progenitor of
SN~2010jl was likely to be a very massive star with an initial mass
above 30 M$_{\odot}$, and with dense CSM expanding at 40--120 km
s$^{-1}$ along our line of sight.  Attributing the IR excess to
pre-existing dust, Andrews et al.\ (2011) infer a very large amount of
mass in the CSM of SN~2010jl, possibly suggesting a massive progenitor
similar to luminous blue variables (LBVs).  Additional information on
the basic observed properties of SN~2010jl can be found in recent
reports (Patat et al.\ 2011; Smith et al.\ 2011a; Andrews et al.\
2011).

\begin{table}\begin{center}\begin{minipage}{3.25in}
      \caption{Spectroscopic observations of SN~2010\lowercase{jl}}
\scriptsize
\tighten
\begin{tabular}{@{}lccccc}\hline\hline
Date &Tel./Inst.$^a$ &Day$^b$ &$\delta\lambda$ (\AA)$^c$ 
   &$\lambda$/$\Delta\lambda$  &$W_{{\rm H}\alpha}$(\AA)$^d$  \\ 
\hline
2010 Nov.\ 5    &K/L &2   &3250--7550  &1100 &150 \\
2010 Nov.\ 5    &M/B &2   &6200--7500  &4500 &133 \\
2010 Nov.\ 6    &M/B &3   &6200--7500  &4500 &141 \\
2010 Nov.\ 7    &K/D &4   &4400--9520  &4400 &188 \\
2010 Nov.\ 15   &L/K &12  &3400--10220 &600  &217 \\
2010 Nov.\ 30   &L/K &27  &3400--10220 &600  &297 \\
2011 Jan.\ 4    &L/K &62  &3400--10220 &600  &519 \\
2011 Jan.\ 16   &M/B &74  &6200--7500  &4500 &586 \\
2011 Jan.\ 26   &L/K &84  &3400--10220 &600  &670 \\
2011 Feb.\ 2    &L/K &91  &3400--10220 &600  &688 \\
2011 Feb.\ 9    &L/K &98  &3400--10220 &600  &758 \\
2011 Feb.\ 23   &M/F &112 &8000--23000 &6000 &... \\
2011 Mar.\ 31   &L/K &148 &3400--10220 &600  &1020 \\
2011 May.\ 12   &L/K &190 &3400--10220 &600  &1120 \\
2011 Jun.\ 27   &M/B &236 &6300--7600  &4500 &1350 \\
\hline \\
\end{tabular}

$^a$Telescope/Instrument: K/L or K/D = W.M.\ Keck Observatory with
LRIS or DEIMOS; M/B = Multiple Mirror Telescope with the Blue Channel 
spectrograph; L/K = Lick Observatory 3m Shane telescope and the Kast 
double spectrograph; M/F =  Magellan Observatory and the FIRE 
spectrograph. \\
$^b$ Days after discovery. \\
$^c$ Rest wavelengths.  \\
$^d$ Emission equivalent width of H$\alpha$.
\label{tab:spectab}
\end{minipage}\end{center}
\end{table}

\begin{figure}
\epsscale{0.95}
\plotone{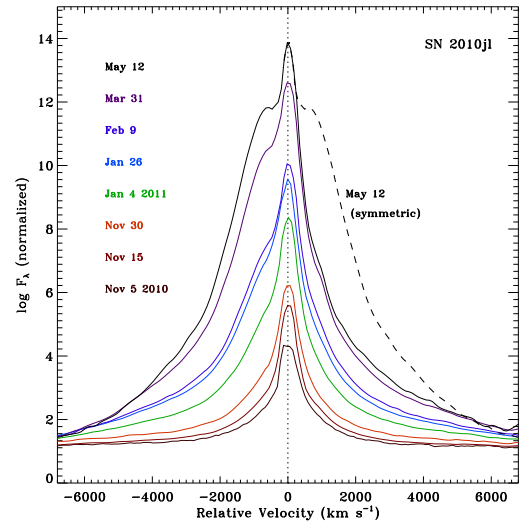}
\caption{Evolution of the H$\alpha$ profile of SN~2010jl, showing that
  the red side of the intermediate-width component weakens
  systematically compared to the blue side during the 6 months after
  discovery.  The dashed curve shows the blue wing of the line
  reflected to the red side, demonstrating what the profile would look
  like if it were symmetric on the final epoch on May 12.}
\label{fig:ha}
\end{figure}

\section{OBSERVATIONS}

\begin{figure}
\epsscale{0.95}
\plotone{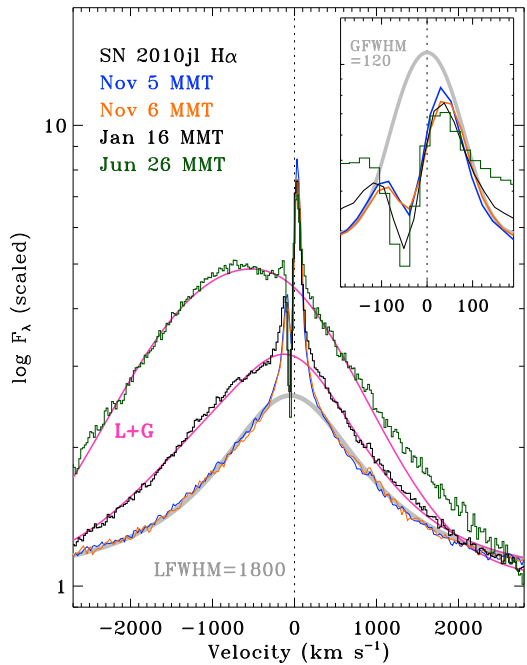}
\caption{High-resolution spectra obtained with the MMT/Blue Channel showing
  the H$\alpha$ line profile of SN~2010jl.  All epochs used the same
  instrument configuration.  This is similar to Figure 3 from Smith
  et al.\ (2011a), except that we have added two additional epochs of
  MMT spectra.  The gray curve shows the same Lorentzian profile as in
  the previous paper, with a Lorentzian full width at half-maximum 
  intensity (LFWHM) of 1800 km s$^{-1}$. The magenta curves show the same
  Lorentzian profile plus a Gaussian shifted by $-$700 km s$^{-1}$ in
  both cases; the Gaussian component is obviously stronger for the
  magenta curve associated with the 2011 June 27 profile.  The inset
  concentrates on the narrow H$\alpha$ component from the CSM; here the
  gray curve is a symmetric Gaussian that is the same as in Smith et
  al.\ (2011a).}
\label{fig:ha2}
\end{figure}

We obtained visual-wavelength spectra of SN~2010jl on several epochs
during the first $\sim$7 months after discovery.  The observations and
data reduction for the two epochs of Keck spectra (2010 Nov.\ 5 and 7)
were already described in our previous paper (Smith et al.\ 2011a),
where we also discussed the observations and data reduction for the
first two epochs of spectra obtained with the Blue Channel spectrograph
on the MMT (2010 Nov.\ 5 and 6).  The high-resolution MMT spectra
obtained on 2011 Jan.\ 16 and June 27 are presented here for the first
time, but the observational setup and data reduction followed the same
procedures as for the 2010 Nov.\ MMT spectra described earlier (Smith et
al.\ 2011a).  We also present 8 additional epochs of optical spectra
obtained with the Kast double spectrograph (Miller \& Stone 1993)
mounted on the Lick Observatory 3~m Shane telescope.  All spectra were
obtained with the long slit oriented at the parallactic angle
(Filippenko 1982) to minimize losses caused by atmospheric dispersion.
The spectral observations are summarized in Table 1.  Most of the
spectra are shown in Figure~\ref{fig:spec} (excluding the
high-resolution MMT spectra which only covered a portion of the
optical wavelength range), while Figures~\ref{fig:ha} and
\ref{fig:ha2} illustrate details of the H$\alpha$ line profile.

\begin{figure}
\epsscale{0.95}
\plotone{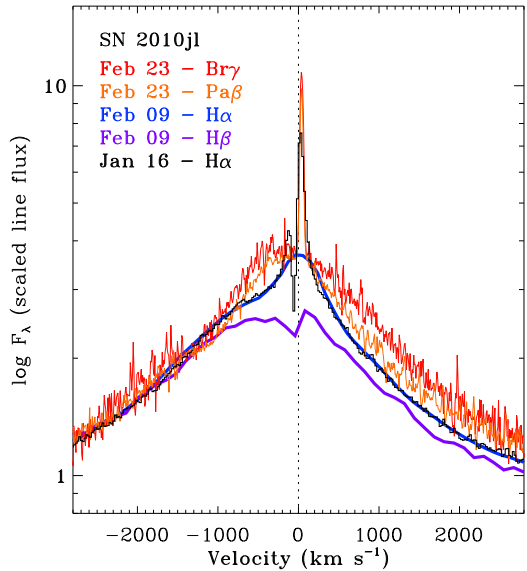}
\caption{Wavelength dependence of the line-profile asymmetry in
  SN~2010jl. The H$\alpha$ profile from the 2011 Jan.\ 16 MMT spectrum
  (black) is the same as in Figure~\ref{fig:ha2}.  The H$\beta$
  (purple) and H$\alpha$ (blue) profiles are from a single spectrum
  taken at Lick on 2011 Feb.\ 9.  The H$\alpha$ profile did not change
  in the $\sim$3 weeks between these observations (the difference at
  line center is due to the different spectral resolution used in the
  MMT and Lick spectra; see Table~\ref{tab:spectab}).  The Pa$\beta$
  (orange) and Br$\gamma$ (red) line profiles are from the FIRE
  spectrum taken 2 weeks later.  These line profiles show that the
  asymmetry is more pronounced at shorter wavelengths, consistent with
  dust being the agent responsible for the blueshift of optical
  lines. The IR line profiles are less asymmetric, despite being taken
  after the optical lines shown here.}
\label{fig:fire}
\end{figure}

We also obtained one near-IR spectrum of SN~2010jl on
2011 Feb.\ 23 using the Folded port IR Echellete (FIRE)
spectrograph\footnote{\tt http://web.mit.edu/~rsimcoe/www/FIRE/.}
mounted on the 6.5~m Baade telescope of the Magellan Observatory.  We
used FIRE in the high-resolution echellete mode with a slit width of
0$\farcs$6, yielding a spectral resolving power of $\sim$6000 or about
30 km s$^{-1}$.  Sky subtraction was accomplished by nodding along the
slit in ABBA sequences, with a total on-source integration time of 20
min.  In this paper, we are primarily interested in comparing the line
profiles of near-IR hydrogen lines to those of visual-wavelength
hydrogen lines that exhibit asymmetry.  Figure~\ref{fig:fire} shows the
line profiles of Pa$\beta$ and Br$\gamma$ in the 2011 Feb.\ 23 FIRE
spectrum, compared to the H$\alpha$ and H$\beta$ profiles obtained a
few weeks earlier.

\section{DISCUSSION}

\subsection{Results}

The main observational results of this paper are that (1) the H$\alpha$
line profile of SN~2010jl provides a textbook example of the
systematic blueshift of emission lines that is normally taken as clear
evidence for dust formation in SNe, and (2) this asymmetry is stronger
at shorter wavelengths as one expects if dust is responsible.
Figure~\ref{fig:ha} shows the evolution of the H$\alpha$ emission line
during the first 6 months after discovery. Like SN~2006jc (Smith et
al.\ 2008a), the systematic blueshift is present in the
intermediate-width component, at velocities within $\pm$4000 km
s$^{-1}$ that arise in the post-shock region of CSM interaction. The
profile is relatively symmetric at first, but as time proceeds the
blue side of the line becomes much stronger than the red side 
(Figure~\ref{fig:ha}). Also like in SN~2006jc, the line asymmetry is 
wavelength dependent (Figure~\ref{fig:fire}).

The H$\alpha$ equivalent width increases steadily during this time,
from 150 \AA \ on 2010 Nov. 5 to 1350 \AA \ on 2011 June 27, and the
maximum velocity of the wings of H$\alpha$ increases from 3000 km
s$^{-1}$ to about 15,000 km s$^{-1}$ during the same time.  (Changes
in the intronsic line flux will be discussed in a future paper.)  The
characteristic continuum temperature around 7,000 K stayed roughly
constant during this time.  Interestingly, the one obvious qualitative
change in the spectrum is in the Ca~{\sc ii} IR triplet around 8500
\AA, which is absent in the earliest epochs, but then strengthens at
later epochs. We return to the possible significance of this below.

Figure~\ref{fig:ha2} shows high-resolution MMT spectra of the
H$\alpha$ profile at three different epochs.  Immediately after
discovery, the intermediate-width component is dominated by a
Lorentzian profile that may indicate electron scattering through high
optical depths in the CSM (see, e.g., Smith et al.\ 2010), as we noted
in our previous paper.  A few months later, however, the
intermediate-width component has qualitatively changed.  The observed
intermediate-width profile can be matched by adding a Gaussian with
FWHM = 1000 km s$^{-1}$ that is offset by $-$750 km s$^{-1}$ (solid
magenta curve), in addition to the same Lorentzian.  The relative
strength of the Gaussian component (with the same width and offset)
increases by the last epoch on 2011 June 27 (Figure~\ref{fig:ha2}).

The evolution of the narrow component of H$\alpha$ arising in the
pre-shock CSM is also interesting, shown in the inset of
Figure~\ref{fig:ha2}.  Although the red emission component appears
relatively constant from Nov.\ 2010 to Jan.\ 2011 (note that these MMT
spectra are taken with the same instrument configuration), the
blueshifted P Cygni component, due to absorption by CSM along the line
of sight, {\it becomes stronger and shifts farther toward the blue}.
A similar effect was seen in the spectral evolution of SN~2006gy
(Smith et al.\ 2010), which was taken as evidence for a Hubble-like
expansion pattern in the CSM (i.e., faster expansion speeds at larger
radii in the CSM).  This, in turn, supported the hypothesis that the
pre-SN CSM of SN~2006gy was formed in an LBV-like outburst rather than
in a steady wind, and a similar scenario may apply to SN~2010jl.  This
appears to be consistent with the very massive progenitor star
inferred for SN~2010jl (Smith et al.\ 2011a) and its dense, dusty CSM
(Andrews et al.\ 2011).  However, additional analysis of the CSM emission
lines is needed.

\subsection{Interpretation}

If the formation of new dust is responsible for the systematic
blueshift of the H$\alpha$ line, then the new dust could reside in
either the post-shock cooling zone or in the SN ejecta, since either
could potentially block the receding side of the CSM interaction shock
that produces the intermediate-width profiles (see Smith et al.\
2008a).  The time of dust formation is very soon after explosion,
beginning after just 30 days.  This is much sooner than the usual
onset of dust formation in SNe, which typically occurs after a few
hundred days.  The hypothesis of post-shock dust formation therefore
seems more likely, as only a dense radiative shock provides a
mechanism for the rapid cooling.


Andrews et al.\ (2011) favor the interpretation that the IR excess
they observe in SN~2010jl is due to an IR echo caused by pre-existing
CSM dust that is warmed by the luminosity of the SN, and not due to
newly formed dust.  They favor this interpretation based on the
relatively cool dust temperatures of $T = 750$~K in their model, whereas
they expect newly formed dust to emit at 1600 K. Andrews et al.\
(2011) did mention, however, that without a detailed study of the
optical emission-line profiles, it is difficult to rule out the
possibility of new dust formation as an explanation for the IR
emission.  As noted above, the H$\alpha$ profile does in fact show
very clear evidence of a systematic blueshift with time, which would
normally be taken as strong evidence for new dust formation.  The
wavelength dependence of the asymmetry in the hydrogen line profiles from
the blue to the IR seems to clinch the hypothesis that newly formed
dust is responsible for the observed line asymmetry.  This, however,
seems to be at odds with the apparent dust temperatures.

High condensation temperatures around 1600~K are indeed appropriate
for carbonaceous grains, as had been seen in SN~2006jc (Smith et al.\
2008a).  SN~2006jc was a SN~Ibn, which might have involved a SN~Ic
expanding into a He-rich shell, so in that case C-rich dust seemed
likely.  However, some other dust species such as typical silicates
have lower condensation temperatures closer to 1000~K, and in SNe~IIn
we might expect O-rich grains like silicates to be more plentiful than
C-rich dust.  A lower assumed condensation temperature might allow the
hypothesis of newly formed dust to be more consistent with
observations.  Moreover, we note that the single dust temperature
model used by Andrews et al.\ somewhat underpredicts the observed
2~$\mu$m flux (see their Figure 3), so perhaps there is room in the
spectral energy distribution for some contribution from dust that is
warmer than 750~K.  Our spectral observations show that much of the
new dust causing asymmetric line profiles had already formed long
before day 90, which is the time of the IR photometry analyzed by
Andrews et al.\ (2011); perhaps some of that newly formed dust had
already cooled by day 90.

These details of the dust temperature aside, it is nevertheless
possible that an IR echo produced much or most of the IR emission, as
suggested by Andrews et al.\ (2011).  While the line profiles
presented here provide strong evidence that new dust formed in
SN~2010jl, we do not know the mass or temperature of that new dust
from the line profiles.  It is therefore difficult from these
observations to decide if (1) new dust is solely responsible for the IR
emission, or if (2) new dust formed and dominates the line-profile
asymmetry, but an IR echo from pre-existing circumstellar dust is
still responsible for the observed IR excess at 3--5 $\mu$m.

Option (1) would require that the newly formed dust cools from 1000~K
to 750~K in about a month.  This would require that most of the grains
form in a brief episode before day 90, and then cool as the shell
continues to expand and the grains come into temperature equilibrium
with the post-shock gas.  Careful study of this scenario is required
before one can infer the expected temperature of newly formed dust.
Option (2) would require that the newly formed dust is either less
massive than the pre-existing dust or somehow inefficient at radiating
compared to the circumstellar dust, perhaps because it is located in
optically thick clumps in the inhomogeneous post-shock cooling zone
(e.g., Wooden et al.\ 1993).  If option (2) is correct, it also
implies that near- and mid-IR observations may not be reliable tracers
of new dust formation in SNe.


\subsection{An Alternative Explanation?}

Since the formation of new dust based on asymmetric line
profiles is seemingly at odds with the conclusion that most of the
IR-emitting dust is pre-existing, it may be useful to consider
alternative explanations for the blueshifted line profiles seen in
SNe~IIn.  The intermediate-width lines in SNe~IIn are produced in a
physical region that is very different from line forming regions in
normal SN ejecta, so careful consideration of possible sources of
asymmetry are warranted.  No alternative explanation has yet been
proposed to account for the systematic blueshift in normal SNe or
SNe~IIn.  We mention one speculative possibility here, which may
follow from the extremely high optical depths and a consequent
transition from optically thick to optically thin emission that occurs
in SNe~IIn (Smith et al.\ 2008b).

At early times, the symmetric Lorentzian line-profile shape seen in
SNe~IIn might not arise from the post-shock gas at all, but instead
from an emission line formed in the slow pre-shock CSM
that is broadened by electron scattering as the H$\alpha$ line photons
diffuse through the dense CSM ahead of the shock (Smith et al.\ 2010;
Chugai 2001).  In other words, the continuum photosphere is actually
located ahead of the shock in the slow CSM, and the observed line
wings do not trace the kinematics of the SN.  Instead, the red wing of
the line is an effect of electron scattering, and it is not directly
related to redshifted gas on the receding side of the SN.

Later on (after a few months in the case of SN~2010jl) the true
intermediate-width component from the expanding post-shock gas is
revealed once the pre-shock CSM becomes optically thin (i.e., the
photosphere recedes back through the forward shock and exposes the
expansion kinematics of the shock for the first time; see Smith et
al.\ 2010).  As the continuum photosphere recedes deeper into the
post-shock zone or into the SN ejecta, it may then reveal H$\alpha$
emission from the near side of the shock --- {\it but the continuum
  photosphere or dust may still block the far side of the post-shock
  shell, leading to the appearance of a blueshifted line profile}.
Interestingly, we note that the time when H$\alpha$ develops
asymmetric blueshifted profiles coincides roughly with the time when
the emission from the IR Ca~{\sc ii} triplet strengthens
(Figure~\ref{fig:spec}).  The onset of asymmetry also occurs when the
line profile changes from Lorentzian (dominated by scattering)
to predominantly Gaussian (Figure~\ref{fig:ha2}).  Since the
characteristic continuum temperature stays roughly constant (such that
the growth in Ca~{\sc ii} is not attributable to a change in
temperature), this suggests that we are seeing deeper into the 
post-shock gas or SN ejecta at roughly the same time when the 
H$\alpha$ line starts to become asymmetric.

In this case, the line profiles are already asymmetric and are
``exposed'' as the continuum scattering photosphere recedes through
the forward shock.  The asymmetric blueshift of the lines could be due
to either extinction by newly formed dust in the post-shock shell, or
to high continuum optical depths.  In the case of
SN~2010jl, we have presented IR spectra demonstrating that the effect
is wavelength dependent, supporting the interpretation that dust is
responsible in this object.  However, other cases of SNe~IIn may exist
where the wavelength dependence is not seen, and high
continuum optical depths may be responsible for line asymmetry. We
note that many SNe~IIn remain luminous in the continuum for long
periods of time after explosion, requiring the presence of high
continuum optical depths.  This ambiguity highlights the importance of
obtaining spectra that span a wide wavelength range to look for
reddening effects.  Detailed radiative transfer simulations to
investigate the feasibility of this effect would be valuable.

One additional test is that at much later times when the material is
assured to be optically thin, the lines may remain asymmetric if dust
is the responsible agent, whereas the profiles should become
symmetric again if the continuum optical depth drops and the material
becomes transparent.  SN~2010jl is so bright that it may afford us the
possibility to study its line profiles at very late times.
Interestingly, we point out that the blueshifted H$\alpha$ line
profile in the SN~IIn 2006tf persisted (and even became more
pronounced) at very late times $\sim$445 days after discovery (Smith
et al.\ 2008b), suggesting perhaps that dust was indeed formed in the
post-shock zones of that SN~IIn as well.

\acknowledgments 
\footnotesize

The work presented here is based in part on observations made at the
the MMT Observatory, a joint facility of the Smithsonian Institution
and the University of Arizona, and at the Lick Observatory, which
is owned and operated by the University of California. 
Some of the data presented herein were
obtained at the W.~M. Keck Observatory, which is operated as a
scientific partnership among the California Institute of Technology,
the University of California, and NASA; the observatory was made
possible by the generous financial support of the W.~M. Keck
Foundation.  We thank the staffs at these observatories for their
efficient assistance, as well as A.\ Barth, S. B. Cenko, J.\ Choi,
A. Diamond-Stanic, R.~J.\ Foley, K.\ Hiner, M.\ Kandrashoff, I.\
Kleiser, A.\ Merritt, J.\ Rex, and J.\ Walsh for their help with
observations at Keck and Lick.  The supernova research of A.V.F.'s group
at U.C. Berkeley is supported by National Science Foundation grant
AST-0908886 and by the TABASGO Foundation. J.M.S.\ is grateful
to Marc J.\ Staley for a Graduate Fellowship.

{\it Facilities:} Keck I (LRIS), Keck II (DEIMOS),
MMT (Blue Channel), Lick (Kast), Magellan (FIRE).


\end{document}